# DPDPU: Data Processing with DPUs

Jiasheng Hu[1], Philip A. Bernstein[2], Jialin Li[3], Qizhen Zhang[1]
[1]University of Toronto, [2]Microsoft Research, [3]National University of Singapore
[1]{jasonhu, qz}@cs.toronto.edu, [2]philbe@microsoft.com, [3]lijl@comp.nus.edu.sg

## ABSTRACT
Improving the performance and reducing the cost of cloud data systems is increasingly challenging. Data processing units (DPUs) are a promising solution, but utilizing them for data processing needs characterizing the new hardware and recognizing their capabilities and constraints. We hence propose DPDPU, a platform for holistically exploiting DPUs to optimize data processing tasks that are critical to performance and cost. It seeks to fill the semantic gap between DPUs and data processing systems and handle DPU heterogeneity with three engines dedicated to compute, networking, and storage. This paper describes our vision, DPDPU's key components, their associated utilization challenges, as well as the current progress and future plans.

## 1 INTRODUCTION

Recent trends in computing and data center architectures have made improving the performance and cost efficiency of cloud data systems increasingly challenging. First, speedup of general-purpose processors is not keeping up with data growth due the slowdown of hardware scaling laws. This has led to degrading compute-bound performance over large volumes of data. Second, high-bandwidth I/O devices, e.g., solid-state drives (SSDs) and network interface cards (NICs), have largely increased the speed of data movement. However, since the CPU instructions incurred per byte access during I/O remains nearly constant [15], moving data at a higher rate consumes significantly more CPU resources. Moreover, cloud providers are evolving their data centers into disaggregated data centers (DDCs). With decoupled compute and data, resource disaggregation intensifies network communication, exacerbating the performance and cost challenges in cloud data systems.

A long line of work has aimed to address these cloud data processing challenges: hardware acceleration using domain-specific hardware (e.g., GPUs [8, 13, 26–28, 38, 43] and FPGAs [11, 18, 39, 48]), OS kernel bypass with userspace I/O (e.g., RDMA [35, 45, 51] and SPDK [1, 15]), as well as caching [55] and compute pushdown [50, 54] to minimize the impact of disaggregation. Each of these proposals has limitations of its own. Hardware acceleration demands deep hardware expertise and embeds domain-specific, non-portable characteristics into system designs; Userspace I/O requires modification to applications for direct hardware access, making it impractical for existing large-scale software systems such as DBMSs to adopt; though effective in improving performance, techniques targeting resource disaggregation fall short to reduce cost. Overall, *there is a lack of holistic platforms that combat cloud data processing performance and cost challenges altogether.*

Data processing units (DPUs) [2–5, 29], the latest generation of programmable NICs (i.e., SmartNICs), emerge as a promising hardware platform. A DPU is a System-on-a-Chip (SoC) equipped with a collection of hardware resources optimized for data-path efficiency. This includes energy-efficient CPUs (e.g., Arm cores), hardware accelerators (e.g., compression and encryption ASICs), network processors, and a moderate amount of onboard memory. DPUs are positioned to overcome the limitations of existing proposals. As an SoC, a DPU runs independently of the host. They can, therefore, augment the overall system architecture without host application modifications, facilitating portability and adoption. In addition, a DPU is capable of offloading host I/O processing at line-rate, reducing resource consumption on the host.

Despite their promises, effective utilization of DPUs for data processing systems has to address the following challenges.
**Challenge #1: abstraction mismatch.** DPUs are packet-oriented networking devices. Consequently, the programming interfaces exposed by DPUs are not intended for data system developers and operators. For instance, NVIDIA's Datacenter-On-a-Chip Architecture (DOCA) [31] and Intel's Infrastructure Programmer Development Kit (IPDK) [16] enable users to build in-network offloading pipelines. They provide libraries such as data plane development kit (DPDK) [40], Open vSwitch (OVS) [41], and P4 [42], with which user programs operate on packets, flows, and raw bytes, rather than data objects (e.g., pages and records). Friendlier DPU interfaces and toolkits are thus needed for data systems.
**Challenge #2: resource diversity.** A DPU SoC consists of a spectrum of hardware resources, ranging from general CPU cores to specialized ASICs for hardware acceleration. Orchestrating these processing units for various data processing tasks and scheduling tasks across them to reflect workload dynamics are non-trivial. A static partitioning scheme, e.g., having the ASICs to execute supported workloads, while the CPUs process the remaining tasks, can lead to suboptimal resource utilization and load imbalance. Prior work has shown that core scheduling on the DPU alone poses challenges [24]. Incorporating other resources such as onboard memory, high-speed network interfaces, and direct access to PCIe peer devices only adds to the allocation and scheduling complexities.
**Challenge #3: DPU heterogeneity.** Hardware vendors and cloud providers are producing their own DPUs, such as NVIDIA BlueField [4], Intel IPU [3], Microsoft Fungible [2], Alibaba CIPU [5], and AWS Nitro [6]. Running data systems on heterogeneous DPUs across different data centers requires an infrastructure that facilitates portability. Even though DPUs share similar high-level architectural characteristics (3), they differ significantly in detailed hardware specifications. For example, NVIDIA BlueField-2 is equipped with a regular expression hardware accelerator, which is missing in Intel IPU, or even NVIDIA's own BlueField-3; BlueField-3 supports generic code offloading to NIC cores, while most other DPUs only support match-action table style network offloading. Making things worse, DPU vendors often provide their own proprietary SDKs for programming the board. Without a portable framework, developers have to manually rewrite all DPU-specific optimizations when switching to a different hardware.

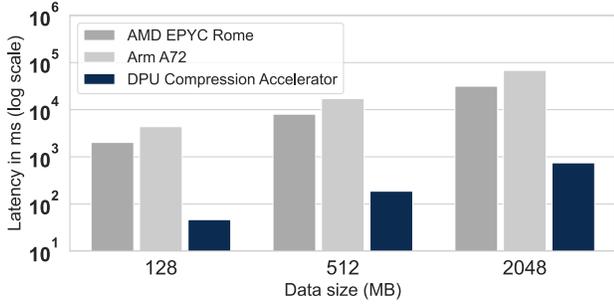

Figure 1: Compression performance on different hardware

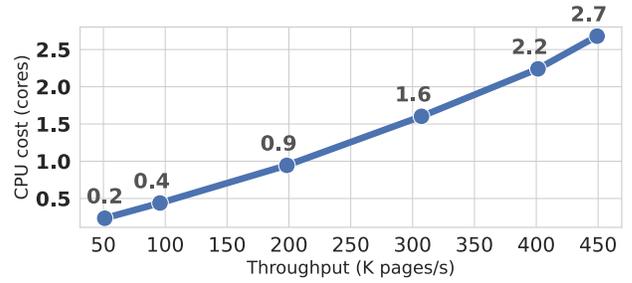

Figure 2: CPU consumption of storage access

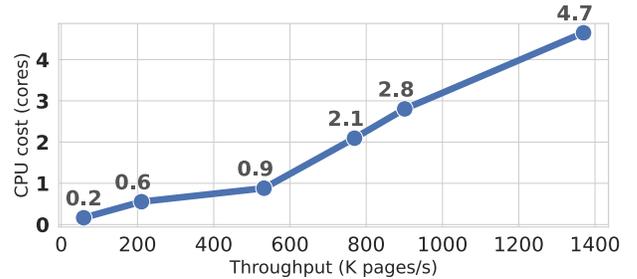

Figure 3: CPU consumption of network communication

This paper proposes DPDPU, a holistic DPU-centric framework for cloud data processing. Our key insight is when the aforementioned impediments are tamed, data systems can efficiently exploit DPUs to optimize a wide spectrum of tasks, e.g., workloads that are compute-intensive, network-intensive, or storage-intensive.

DPDPU includes three components to judiciously harness the various DPU resources: *a compute engine* that runs on DPU CPU cores and hardware accelerators for computational tasks such as expensive on-path data operations (e.g., compression and encryption) and pushdown database operators (e.g., predicates and aggregation); *a network engine* that offloads communication primitives from host CPUs to the DPU network interfaces; and finally, *a storage engine* that leverages direct storage device access to improve local and disaggregated storage performance while saving cost. DPDPU further schedules tasks across DPU hardware accelerators, DPU CPUs, and host CPUs based on task specifications and resource availability.

DPDPU offers high-level, hardware-neutral interfaces to ease programming and porting effort for DPU accelerated data systems. Specifically, users write stored procedures to express tasks in the compute engine. The network and storage engines expose a familiar asynchronous I/O abstraction, allowing existing data systems to adopt DPDPU with minimal effort. We eschew vendor-specific features in the framework such that customized optimizations atop DPDPU are portable across different DPUs. To handle hardware heterogeneity, we propose a *DP kernel* abstraction that unifies hardware accelerators and CPUs. When a DP kernel is not supported by any accelerator, DPDPU executes it on DPU CPUs or host CPUs and inform the decision to the application.

This paper makes the following contributions.
- We demonstrate the performance and cost challenges in cloud data processing (Section 2), and show the opportunities enabled by DPUs (Section 3).
- We present the overall vision of DPDPU (Section 4).
- We discuss challenges in each DPDPU component and propose the high-level design (Sections 5, 6, and 7).
- We survey related prior work (Section 8), report the current progress, and propose the next steps for DPDPU (Section 9).

## 2 EMERGING CHALLENGES IN THE CLOUD

In this section, we demonstrate the performance and cost efficiency challenges of running data systems in the cloud. We provide quantitative evidences using micro-benchmarks, where we measure the efficiency of compute- and I/O-intensive tasks as well as the overhead of resource (storage) disaggregation.

### 2.1 Compute Inefficiency

It is well-known that CPU speed-ups have slowed down over the past decade. On the other hand, data systems frequently invoke compute-heavy subroutines. For instance, DBMSs often compress and encrypt data before network transfers to reduce network traffic and to provide data privacy and security. Could data systems still rely on CPUs to sustain good performance on these compute tasks?

To answer this question, we measured the performance of data compression (with the lossless DEFLATE algorithm [33]) on natural language datasets of various sizes on an AMD EPYC CPU and an Arm CPU. Figure 1 shows that, while the more advanced EPYC CPU outperforms the Arm CPU, they both suffer from high and growing latency when more data is compressed. This result shows that it is intractable for data systems to perform compute-intensive operations when managing large-scale data.

### 2.2 I/O Cost

We now measure the CPU cost of performing high-bandwidth I/O, which is among the most common tasks in database systems. Specifically, we evaluate the CPU consumption of accessing 8 KB pages from Linux-managed SSDs.

As we observe in Figure 2, the number of CPU cycles grows linearly with increasing I/O throughput. When the throughput reaches 450 thousand pages per second, the average CPU consumption is as high as 2.7 cores. We also tested Linux storage performance with the more recent `io_uring`, but observed similar CPU cost. The experiment here demonstrates that data systems with high I/O requirements will consume significant CPU resources, which translates to higher hardware cost.

## 2.3 Disaggregation Overhead

Lastly, we assess the overhead of resource disaggregation. In particular, storage disaggregation, where compute and storage are hosted on different servers connected via the network, has been commonplace in today's cloud data centers. The architecture enables better flexibility in resource management, but at the expense of additional network I/O for storage accesses. leading to higher access latency and even more CPU consumption.

To quantify the latency and CPU consumption overhead of disaggregation, we measure the cost of network communication via TCP/IP sockets for transferring 8 KB pages over a 100 Gbps network. As shown in Figure 3, the additional network I/O induced by disaggregation consumes significant CPU resources, particularly at higher bandwidth. Note that such I/O processing are competing with other compute tasks, such as those in Section 2.1, for CPUs.

## 3 DPU OPPORTUNITIES

DPUs are SoC-based SmartNICs[1]. Figure 4 shows the architecture of NVIDIA BlueField-2 (BF-2) [4], a popular DPU in mass production. Resources on a DPU can be categorized into five types: (1) energy-efficient CPU cores, (2) onboard memory, (3) hardware accelerators, (4) high-speed network interfaces, and (5) PCIe interface. Specifically, BF-2 consists of 8 Arm A72 cores with a clock rate of 2.5 GHz, 16 GB DDR 4 memory, a set of hardware accelerators that includes regular expression, compression, encryption, and deduplication, ConnectX-6 NIC with 100 Gbps bandwidth, and a PCIe 4.0 switch that have access to host memory and other PCIe-connected devices, such as SSDs and GPUs. Although the detailed hardware configuration varies across different vendors, this characterization can be generalized to other DPUs, e.g., Intel IPU [3] and Microsoft Fungible [2]. These resources, combined with DPU data-path optimizations, can be leveraged to address the above challenges.

Specifically, to improve compute efficiency, data systems can utilize the hardware accelerators to execute compute-intensive operations on the data path. These accelerators are ASICs designed for specific compute tasks; specialization improves power efficiency and performance compared to general-purpose processors. Figure 1 shows that the compression accelerator on BF-2 outperforms CPUs by an order of magnitude. To reduce I/O cost, DPUs usually provide advanced userspace libraries to build efficient I/O pipelines. For instance, BF-2 employs SPDK and DPDK for users to directly access storage devices and the network interface in the userspace without involving the host. Together with the general-purpose CPU cores and moderate amount of memory on the DPU, users can build arbitrary, low-latency, and high-bandwidth I/O services to free the host from expensive storage and network I/O activities.

Despite the potential benefits, challenges around abstraction mismatch, resource diversity, and DPU heterogeneity must be addressed in order to better utilize DPUs for data processing systems.

## 4 THE DPDPU FRAMEWORK

We envision a DPU-powered software platform that exploits the opportunities in Section 3 to tackle the cloud data processing problems in Section 2. The platform, which we term DPDPU, does so by (1) bridging the semantic gap between raw DPU resources and cloud data processing tasks, (2) efficiently utilizing diverse hardware resources on individual DPUs, and (3) decoupling the detailed hardware configurations among different DPUs from the optimizations at the data system layer. As shown in Figure 5, DPDPU consists of three modules that allow for optimizing compute-intensive and I/O-intensive operations.

---

[1]The other major category is FPGA-based SmartNICs. We focus on SoC-based SmartNICs for their easier programmability and development process.

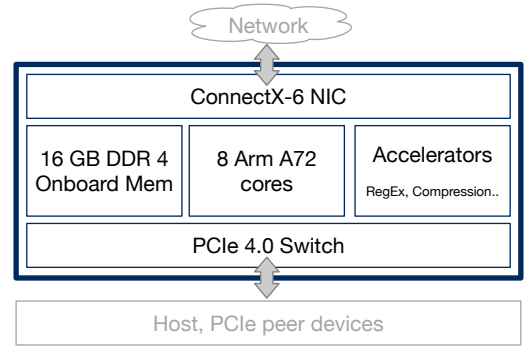

Figure 4: NVIDIA BlueField-2 DPU architecture

**Components and accessed resources.** We now describe the DPU resources managed by each component and the interactions between components. The next sections discuss detailed designs of each component and the key challenges.

Compute Engine offers *efficient* and *versatile* computational power for data processing tasks. The engine carefully orchestrates compute-intensive tasks across four types of compute resources: DPU onboard CPUs, DPU hardware accelerators, host CPUs, and other popular data center accelerators, e.g., GPUs and FPGAs, connected via PCIe. The working set of execution can be cached in both DPU memory and host memory.

Network Engine handles network I/O. It utilizes the advanced networking facilities built in DPUs (high-speed interfaces, match-action offloading, and user libraries) to improve network I/O efficiency. More specifically, the DPU DMA engine serves as an abstraction boundary to decouple the front end of popular networking approaches (utilized by host applications) from their protocol execution, which is offloaded to the DPU using onboard memory, CPU, and the network interface.

Storage Engine improves storage path efficiency, including requests from both local applications and those from remote clients. For local applications, the engine offers a light-weight user library to forward storage requests from the client to the DPU, where it accesses SSDs via PCIe peer-to-peer communication. For requests from remote clients, it coordinates with the Network Engine to execute storage requests immediately on the DPU without involving the host.

**Interactions.** The different DPDPU components can be *composed* to execute complex tasks in data systems. For instance, in response to a remote storage access request with compressed data, a DPDPU program may first read the data from local SSDs using the Storage Engine. It then invokes the Compute Engine to compresses the data in the DPU compression accelerator. Finally, the Network Engine delivers the result to the client.

Take predicate pushdown as another example. Leveraging DPDPU, the storage server first reads the database records from SSDs through the Storage Engine. It then directly applies predicates on these tuples using the Compute Engine, and only sends the qualified tuples back to the remote database server via the Network Engine.

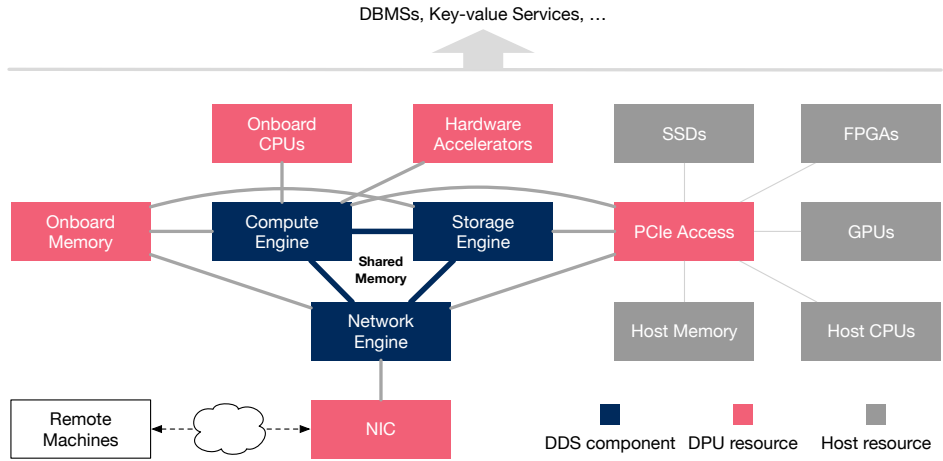

Figure 5: DPDPU components and resources they access

DPDPU facilitates composability using two mechanisms. First, it enables shared state across the three engines via the DPU memory. The schema of the state and cached data are customizable by the application. Note that within each component, consistency is not guaranteed due to asynchronous accesses, e.g., the network, the hardware accelerators, and host resources via PCIe.

Moreover, DPDPU enables efficient, streamlined data communication across engine boundaries. To do so, the API and the execution model of the engines facilitate pipelined data processing—one engine's output can be streamed to another engine without waiting for the completion of work in progress. This allows for constructing efficient asynchronous pipelines that overlap I/O and computation.

## 5 COMPUTING

We design the Compute Engine (CE) with the following goals.

(1) **Efficient.** As the primary motivation for CE is to address compute inefficiency, we aim to maximize the efficiency of compute tasks that are offloaded to CE.
(2) **General-purpose.** To benefit various cloud data processing systems, the CE should handle a wide spectrum of tasks, from data-path primitives (e.g., compression and encryption) to relational operator pushdown.
(3) **Easy to program.** A major difficulty of programming DPUs is the low-level interfaces across different processing units. CE offers APIs already familiar to the data system developers to improve usability.
(4) **Portable.** In addition to portability across DPUs, CE must also account for the diverse compute resources on the DPU and the host when executing the same user tasks.

**Interface.** We provide stored procedures (sprocs) for users to express their compute tasks. Previous work [52] has explored using sprocs as a general programming abstraction to offload computation for data processing systems. Despite their benefits, sprocs are primarily designed for CPU execution; the abstraction lacks native support for hardware acceleration. To overcome this limitation, we introduce *DP kernels*, an extensible set of specialized functions built in DPDPU that optimizes sproc execution efficiency. The user can query what DP kernels are available in the CE and select the one that matches the application need. DP kernels, however, do not expose hardware-level details to developers. Sprocs with DP kernels naturally satisfy CE's general-purpose and easy to program goals. We next discuss execution details and explain how CE achieves efficiency and portability.

**Execution.** A sproc is first registered with the CE, which pre-compiles it into a shared library. At runtime, the library is loaded into the user program and runs on a DPU CPU core. DP kernels, on the other hand, represent compute-heavy tasks and thus are prioritized for hardware acceleration to maximize compute efficiency. However, due to hardware heterogeneity, certain hardware accelerators are not universal across DPU targets. For instance, BlueFiled-2 includes an RegEx engine, which is not available on BlueField-3 and Intel IPU. To ensure that the same user code can run on different DPUs, DP kernels must be portable and backward- and forward-compatible.

To that end, we require that each DP kernel can be executed on *any compute hardware*, e.g., CPUs, ASICs, FPGAs, or GPUs. The actual execution during runtime depends purely on hardware availability. We allow the user to specify where a DP kernel is executed (specified execution); alternatively, the CE can construct a schedule for all the DP kernels (scheduled execution). Scheduled execution enables the CE to optimize the overall performance of a sproc given hardware constraints of the target platform.

**Example.** An example of a sproc with a DP kernel is shown in Figure 6. The sproc serves a request from a remote client that reads a set of pages, compresses them, and sends the compressed pages back to the client. Since compression is the most compute-intensive task in this sproc, we accelerate it using the compression kernel (`dpk_compress`). Here, the user first specifies the kernel to be executed on the compression accelerator (line 20). If the accelerator is currently unavailable on the DPU, the user moves the computation to a DPU CPU core (line 24).

Alternatively, the implementation can leave target device unspecified in `dpk_compress`. The kernel will then be scheduled by CE, and the call always returns a valid work item in progress. The

```python
import dpdpu.compute_engine as ce
import dpdpu.network_engine as ne
import dpdpu.storage_engine as se

read_compress_send_pages(req):
  page_read_list = {}
  page_comp_list = {}
  page_send_list = {}
  dpk_compress = ce.get_dpk("compress")

  for net_req in req.pages:
    # async read
    read_req = se.read(net_req.file_id,
      net_req.addr, PAGE_SIZE)
    page_read_list.add(read_req)

  for read_req in page_read_list:
    wait(read_req)
    # async compression (fast)
    comp_req = dpk_compress(read_req.data,
      "dpu_asic")
    if comp_req is None:
      # async compression (slow)
      comp_req = dpk_compress(read_req.data,
        "dpu_cpu")
    page_comp_list.add(comp_req)

  for comp_req in page_comp_list:
    wait(comp_req):
    # async send with TCP
    send_req = ne.tcp.send(req.client,
      comp_req.data)
    page_send_list.add(send_req)

  for send_req in page_send_list:
    wait(send_req)
```

Figure 6: An example of sproc with DP kernels where page compression is accelerated (specified execution). Different modes of execution facilitate portability.

main benefit of specified execution is predictable program behavior; it, however, leaves the burden of optimizing sproc performance to the user.

**Open challenges.** Developing the CE must address several technical challenges. First, a sproc may be invoked in parallel at a high rate, e.g., upon receiving a packet. As such, *proper scheduling is critical to the overall performance*. Prior work adopted various scheduling disciplines to achieve high NIC offloading performance. For instance, iPipe [24] utilizes a first-come-first-served (FCFS) queue and a deficit round robin (DRR) queue to schedule tasks with low and high variance respectively across DPU CPU cores and host CPU cores. The CE needs to schedule not only sprocs between DPU and host CPUs, but also DP kernels across all computing units. Hardware accelerators exhibit vendor-specific characteristics, (e.g., high throughput with high latency) that are distinct from CPUs. Consequently, the problem space for scheduling in DPDPU is expanded: How to schedule DP kernels on the same accelerator? How to co-schedule sprocs and DP kernels? How to cater for performance targets from different applications?

Second, a server equipped with a DPU can run multiple applications. *It is vital to provide fairness and performance isolation in a multi-tenant setting*. A naive approach can use containers to slice CPUs and memory on both the DPU and the host. A complete solution, however, must also consider hardware accelerators. Compared to CPUs, the accelerator capacities (i.e., the number of concurrent tasks) vary greatly across hardware; there is also a lack of virtualization support on these accelerators. Hence, multiplexing resources and isolate the execution of DP kernels on accelerators present a challenge.

Finally, DPDPU CE can be further augmented when additional common data center accelerators such as FPGAs and GPUs are connected via PCIe. We first need to map DP kernels to these devices and develop efficient data movement plans based on how a sproc and its DP kernels are spread across different locations. Since such accelerators have higher resource capacities (more cores and memory) than that of the DPU hardware accelerators, it makes sense to fuse multiple DP kernels inside the accelerator to minimize execution latency. In addition, we need to extend the solutions to the previous challenges to incorporate more PCIe devices.

## 6 NETWORKING

Our primary goal for the Network Engine (NE) is to lower communication overhead while maintaining high performance for popular transport protocols, e.g., TCP and, more recently, RDMA. The principle in designing NE is to offload CPU consuming network activities to the DPU, while leaving only light-weight front-end libraries that emulate existing communication frameworks' APIs. This is enabled by DPU's DMA and packet generation capabilities.

**Optimizing TCP.** The traditional TCP/IP stack remains the most popular protocol for network communication in data processing systems. As shown in Section 2, high TCP throughput consumes substantial host CPU cycles. Recent proposals seek to improve TCP CPU efficiency by partial offloading of the TCP stack to DPUs. For instance, IO-TCP [20] divides TCP into a control plane (connection management, congestion control, etc.) and a data plane (data transmission); it runs the former on a single host CPU core, and the latter on the DPU. These solutions, however, target specific applications (e.g., IO-TCP for streaming media files) and require application modifications.

To support general communication for distributed and disaggregated data processing, we propose to move the TCP/IP stack to the DPU and provide a POSIX-like socket API for host applications through a user library. Doing so requires tackling two challenges. First, as the CPU on the DPU is significantly weaker than that on the host, the TCP/IP stack on the DPU must be carefully optimized to avoid performance degradation. Second, as network messages are eventually processed on the host, flow control now spans the host and the DPU. We must co-design TCP on the DPU and host-DPU communication to reflect the signals from host applications in the flow control protocol.

**Optimizing RDMA.** Remote Direct Memory Access (RDMA) has emerged as a promising data center networking technology for achieving high-speed network communication in data processing systems [35, 51]. RDMA runs in userspace and can completely bypass OS overheads. It can also eliminate remote CPU involvement via direct memory access from NIC hardware. To best utilize RDMA for database systems, DFI [45] layers a data flow interface atop the transport to provide pipelined, thread-centric flow execution. It achieves communication performance that is close to the raw RDMA network.

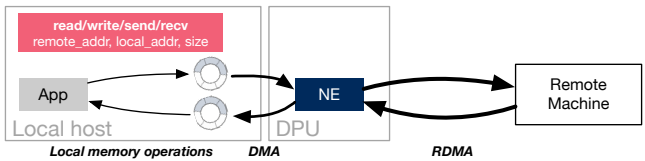

Figure 7: DPU-optimized RDMA

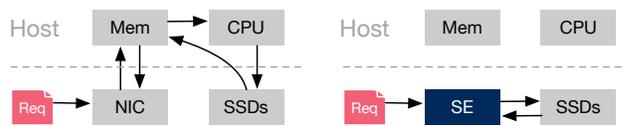

Figure 8: Round trips from NIC to host in today's disaggregated storage (left) can be saved with DPDPU SE (right).

Despite its performance benefits compared to traditional networking stacks, issuing RDMA operations is still CPU costly. For instance, accessing the send/receive queues in a RDMA queue pair requires spinlocks and memory fences to ensure queue ordering. CPU stalls can also happen when ringing the doorbell register of the RDMA NIC. These overheads have been confirmed by recent work [10].

Figure 7 depicts our proposal for optimizing RDMA communication. The design offloads the heavy issuing-side RDMA handling to the DPU. We first replace the RDMA queues with lock-free ring buffers to accept user requests. These buffers are DMA-accessible such that NE on the DPU can poll user requests using the DPU DMA engine. Upon receiving requests, NE issues corresponding RDMA read/write or send/receive to access memory on the remote machine. This asynchronous execution of RDMA must be served together with the non-blocking interface on the host, such that applications only spend minimal resources polling responses.

Cowbird [10] proposes an asynchronous I/O abstraction for disaggregated memory; it offloads RDMA to programmable switches and harvested VMs. DPDPU NE can be viewed as an extension to Cowbird that targets general network communication, supporting both one-sided and two-sided RDMA. The key challenge is to co-design the interface and the execution of the complete set of RDMA operations such that resource consumption on both the host and the DPU can be minimized.

We can apply NE to optimize DFI. Specifically, DFI's interface and its RDMA execution can be decoupled such that data systems running on the host still send records to remote machines using the flow interface. These requests are cached on the host memory and then moved to the DPU for further data flow processing. Doing so requires redesigning DFI's RDMA-accessible buffers with host-managed DMA buffers and DPU-managed RDMA buffers.

## 7 STORAGE

The Storage Engine (SE) in DPDPU is motivated by two advantages of DPUs for storage: first, offloading file-related operations on to the DPU can free significant host resources; second, the DPU sits on the data path to serve requests for disaggregated storage. The former is apparent given the high CPU consumption of storage I/O, as showed in Section 2. Figure 8 demonstrates the latter: when a remote storage request arrives at the DPU, SE can service the request immediately by accessing PCIe-connected SSDs. In comparison, existing disaggregated storage must process the request using host CPUs, incurring additional PCIe, OS, and storage stack overheads.

**Offloading file execution.** We first propose a DPU-backed storage framework that offers a POSIX-like file system API for host applications to manage files and perform file I/O. The processing of file requests is offloaded to the DPU, where we build a *file service* leveraging userspace storage solutions, e.g., Storage Performance Development Kit (SPDK), to optimize file I/O efficiency. Similar to NE, the contention between application threads for issuing file requests and polling responses is minimized with lock-free ring buffers in the user library, and the requests are lazily DMA'ed by the DPU.

Our design requires delegating the management of SSDs from host servers to DPUs, which is a popular trend of adopting DPUs [32].

**Offloading remote requests.** To fully exploit DPU for disaggregated storage, we propose an offload-engine in the SE that allows users to directly process remote storage requests on the DPU. Specifically, users supply a UDF that parses network messages to identify remote storage requests that can be offloaded, and translates them into file operations. A simple UDF can extract file ID, offset, size, and I/O type (read or write) from a request and construct the corresponding file operation. Since the DPU already maintains the mapping between user files and physical blocks on the SSDs (i.e., the file mapping) in the aforementioned DPU file service, SE can directly execute the file operation without contacting the host.

**The key challenge** in realizing this design is the limited resources on DPUs. For instance, in cloud-native DBMS architectures, transaction updates are reflected on disaggregated storage servers with log replay [7, 47]. Running the log protocol can consume up to 100s GB memory to cache hot pages to prevent write amplification. This memory footprint is an order of magnitude larger than DPU memory capacity (e.g., 16 GB). Hence, some of the storage requests that are not suitable for DPU offloading must still be forwarded to the host. This partial offloading raises several technical questions: which requests should be offloaded? What should the offloading API look like to reflect the division? How do we split network traffic without violating transport protocol semantics?

## 8 RELATED WORK

SmartNICs and DPUs have been explored for distributed systems and computer networking [9, 12, 14, 17, 19, 21, 23–25, 30, 32, 34, 36, 37, 46, 49]. Below we summarize several recent works in this line.

LineFS [19] improves the efficiency of distributed file system by offloading CPU-intensive tasks to the DPU and use pipeline parallelism to improve performance. Xenic [36] caches data on DPUs to accelerate distributed transactions. hKVS [9] is a KV store that uses DPU memory to cache hot records and meticulously synchronizes updates to the host. iPipe [24] proposes an actor-based execution framework that utilizes DPUs for distributed applications. It enables scheduling and flexible load migration between the DPU and the host. More recently, IO-TCP [20] proposes a host-DPU codesigned TCP that leverages the data-path efficiency of DPUs to

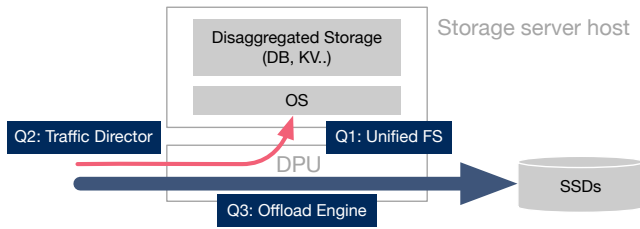
Figure 9: DDS overview

offload the streaming of media files. Lovelock [32] is a DPU-based cluster manager that eliminates the need for host servers, e.g., for hardware accelerators and storage devices.

DPUs are increasingly attracting attention in the database community. Thostrup et al. [44] evaluate the performance benefits of a specific DPU (NVIDIA BlueField-2) for two specific DBMS components (a B-tree index and a sequencer). Their results are aligned with our DPU characterization. SmartShuffle [22] offloads to DPUs low-level networking components as well as DBMS-level tasks in data shuffling.

Differentiating this work from others is the generality of our proposal. It systematically exploits the capabilities of DPU SoCs to tackle a spectrum of challenges in cloud data processing. The three complementary engines in DPDPU present an easily-utilized and portable offering of DPU resources for data system optimizations.

## 9 PROGRESS AND NEXT STEPS

Our first step towards realizing the vision of DPDPU is developing DDS [53], a DPU-optimized disaggregated storage server architecture as part of the Storage Engine. Recall from Section 7 that DPU is inappropriate for fully offloading disaggregated storage requests. Hence, the design of DDS is centered around partial offloading, i.e., remote storage requests are split between the DPU and the host. Specifically, we addressed three key questions: (1) *Q1: how to access files on SSDs directly from the DPU?*, (2) *Q2: how to direct traffic between the DPU and the host?* and (3) *Q3: how to enable general and efficient DPU offloading?*

Figure 9 sketches DDS. To answer the first question, we developed a unified file system that directs file operations on the host to the DPU. Doing so allows the DPU to own the file mapping and thus knows how to serve a remote request. The second question is handled with a traffic director that determines whether each packet should be forwarded to DDS on the DPU or the endpoint on the host. It accomplishes the task without breaking end-to-end transport semantics. Finally, we introduce a high-level API in the offload engine for users to implement the UDF in Section 7, and extensively employ zero-copy to maximize the efficiency of request execution. We integrated DDS with FASTER (a KV store) and Azure SQL Hyperscale (a cloud-native DBMS), two production systems at Microsoft. Empirical studies show that DDS can save up to 10s of CPU cores per storage server.

DPDPU opens a broad space of systems and optimization research for cloud data processing. Our next steps are as follows.

**Caching in DPU-backed file system.** DDS currently achieves minimal memory footprint and has no support for caching on either the host or the DPU in the file system. Provided access to more memory, we can cache hot and warm data to further improve file performance. How to cache, however, is non-trivial because of separate sources of access: caching in host memory is most efficient for host applications, while caching in DPU memory works better for remote requests that can be offloaded. Sizing the cache at the right granularity on the DPU and on the host based on workload characteristics to maximize caching benefit and minimize memory consumption is hence a key challenge.

**Faster persistence.** Techniques such as caching and DDS have been adopted to improve the read query performance for many data systems. Although in cloud data systems, writes are less prevalent than reads, optimizing persistent updates, particularly their end-to-end latencies, is meaningful to mission critical applications and presents unique challenges. More specifically, persistent operations often need to traversing deeper storage stacks than reads; the backing store typically runs on slow hard drives, many even located remotely in disaggregated storage. DPDPU offers opportunities to accelerate persistence performance. By directly connecting DPUs with fast persistent storage (e.g., NVMe SSDs) through PCIe P2P, DPDPU can persist a write request to storage devices or DPU's onboard fast storage before forwarding the operation to the host. Once persisted, the DPU can immediately acknowledge the request without waiting for completion on the host. To realize this design, we plan to design a generic DPU fast-persistence interface. The interface allows various existing data systems to easily benefit from fast persistence with minimum code modifications. We also need to address the challenge of coordinated recovery in this new model, as well as consistency issues arose due to concurrent reads, including both reads forwarded to the host and those offloaded to the DPU, and fast persistent writes.

**Implementing and scheduling DP kernels.** DP kernels are at the core of DPDPU's Compute Engine for harvesting the compute efficiency of various DPU processing units. As detailed in Section 5, designing and implementing these primitives is challenging. The level of abstraction determines whether these functions can capture the requirements of data processing systems and whether they can exploit hardware efficiency. As DP kernels are portable across DPUs, we need to investigate a collection of vendor-provided DPU SDKs, seeking plans that avoid excessive engineering effort.

Another critical task in developing the Compute Engine is to schedule DP kernels (and co-schedule them with sprocs) based on performance requirements of data processing systems.

**Network Engine and database communication optimization.** In addition to the design challenges in Section 6, developing the Network Engine requires mapping out the detailed architectures of our target networking protocols (i.e., TCP and RDMA) and construct the set of cross-host-DPU operations that enables the decoupling of interface and protocol execution.

In our experience, the internal networking stack of cloud-native production DBMSs is a primary source of I/O overhead. We thus plan to dissect the networking stacks of open-source systems to search for a common set of DBMS-specific communication tasks suitable for DPU offloading.